# Structural and Optical Properties of Pulse Laser Deposited $Ag_2O$ Thin Films

Souvik Agasti[*], Avijit Dewasi and Anirban Mitra

*High Power Laser Lab, Department of Physics, Indian Institute of Technology Roorkee, Roorkee-247667, Uttarakhand, India*

[*]*Corresponding author: nandi.agasti@gmail.com*

**Abstract.** We deposited $Ag_2O$ films in PLD system on glass substrate for a fixed partial oxygen gas pressure (70 mili Torr) with the variation of laser energy from 75 to 215 mJ/Pulse. The XRD patterns confirm that the films have well crystallinity and deposited as hexagonal lattice and the crystalline size increases from 26.38 nm to 27.27 nm. The FESEM images show that the particle size of the films increases from 34.84 nm to 65.83 nm. The composition of the films is analyzed from EDX spectra which show that the percentage of oxygen increases from 41.03% to 48.38% with the increment of laser energy. From the optical characterization, it is observed that the optical band gap appears in the visible optical range in an increasing order from 0.87 to 0.98 eV with the increment of laser energy. Our analysis concludes that the $Ag_2O$ thin films, deposited with these parameters, can be considered as a good absorbent layer for solar photovoltaic application.

## INTRODUCTION

Nanostructure thin films have become interesting candidates nowadays to be used in electronic devices. The reduction of dimension introduces quantum properties in these materials and as a result, we see ballistic characteristics in their performance, which makes the device industrially applicable [1]. Silver oxide thin films have novel applications in optical read-write non-volatile memories which can be used for storage applications and it has the property of surface enhanced Raman scattering in plasmonic devices [2,3]. Silver has d-shell electrons which causes different oxidation states: $AgO$, $Ag_2O$, $Ag_2O_3$, $Ag_3O$. The formation of oxide changes with the change of deposition parameters like oxygen pressure, temperature, required energy for oxidation, etc. Silver oxide films has been deposited using many techniques: thermal evaporation [4], RF and DC sputtering [5,6], chemical synthesis [2] and pulsed laser deposition (PLD) [7]. The last one has the advantage of better controllability on oxygen pressure, substrate temperature, excitation energy and moreover cost effective over others, which motivated us to use this technique.

Silver oxide composite has been used as an insulator for coating purpose due to its high band gap [8]. The band gap of electrodeposited $Ag_2O$ nanoparticles has also been reported as 1.46 eV and therefore it has been used as a semiconductor for the photovoltaic purpose [9]. Furthermore, ref. [7] mentions silver oxide films have been deposited using PLD technique which indicates about the possibility of the deposition of $Ag_2O$ film at room temperature as a p-type semiconductor with very low band gap for photovoltaic uses. In this paper, we are presenting four films prepared at room temperature; with a fixed oxygen pressure of 70 mili-Torr and with the variation of laser energy from 75 to 215 mJ/Pulse. The prepared thin films have been characterized using X-ray diffraction (XRD), field emission scanning electron microscope (FE-SEM) and spectroscopic ellipsometry systems.

# EXPERIMENTAL

## Deposition of $Ag_2O$ Thin Films

We prepare $Ag_2O$ thin films on soda lime glass substrates in a PLD system in presence of partial oxygen gas environment at room temperature (~300K). The glass substrates are cleaned with acetone and placed in the PLD chamber. The chamber is first evacuated to the order of $10^{-6}$ Torr. High purity silver (99.99%) target of diameter 3 cm is used for the $Ag_2O$ thin film deposition. Nd:YAG laser operated at the wavelength of 355nm with 9 ns pulse duration is used for this deposition. The substrate to target distance is kept at 5 cm. Four samples are prepared at laser energy 75 mJ/pulse, 130 mJ/pulse, 170 mJ/pulse and 215 mJ/pulse respectively having fixed no. of shots (6000) and a constant partial oxygen pressure (70mTorr) during the deposition.

We observe the oxidation of silver can be changed from $Ag_2O$ to AgO in the PLD systems with the increment of oxygen pressure and also the reduction of oxygen pressure might ensure deposition of Ag only without being oxidized. Indeed, the increment of oxygen pressure ensures more availability of oxygen, which leads to get AgO and the reduction of pressure reduces the availability of oxygen which helps to get Ag instead of $Ag_2O$. This constrains us to concentrate on oxygen pressure at 70 mT in order to get $Ag_2O$ films [7]. The magnitude of base vacuum is very small compared to the oxygen partial pressure and hence, the tiny variation of the base vacuum hardly effects on deposition. The laser energy of higher intensity helps Ag to be excited more. Hence, the no. of plasma ions increase and also the energy of the generated plasma increase and therefore, more probability to interact Ag with oxygen. But in the other way, more exited Ag has more velocity to travel from the target to substrate and therefore gets less time to be oxidized.

## Characterization of $Ag_2O$ Thin Films

The crystal structure characterization is done by X-ray diffraction measurements with the help of Bruker AXS D8 diffractometer advanced in θ-2θ mode. Cu Kα radiation of wavelength 0.154 nm is used for diffraction experiments. XRD data has been plotted and analyzed by the X'Pert Highscore software and the peaks of the samples are matched with literature for identification. The surface morphology of the deposited thin films is examined by Field Emission Scanning Electron Microscope (Carl Zeiss Ultra Plus). The elemental analysis of samples has been performed using energy dispersive X-ray (EDX) spectroscopy. The optical characterization is done in a spectroscopic ellipsometer system (J. A. Woollam Co. Inc.; Model: M-2000.)

# RESULTS AND DISCUSSIONS

The XRD plots of the samples are shown in the Fig. 1. After analysis, it is seen that $Ag_2O$ has been deposited along with the presence of other oxides. But, $Ag_2O$ has a sharper peak of hexagonal crystal lattice along the orientation (100) which is more prominent compared to others. The quality of XRD peaks, depicts the quality of crystal structure; more intense peak indicates better crystallinity. In hexagonal lattice the interlayer distance is related to the lattice parameters by the formula bellow [10].

$$\frac{1}{d^2} = \frac{4(h^2+hk+k^2)}{3a^2} + \frac{l^2}{c^2} \qquad (1)$$

where $d_{hkl}$ is the interplanar distance, calculated using Bragg's law: $2d \sin(\theta) = n\lambda$, for corresponding lattice plane indexed by h, k, l, (they are all integers) and a and c are lattice constants.

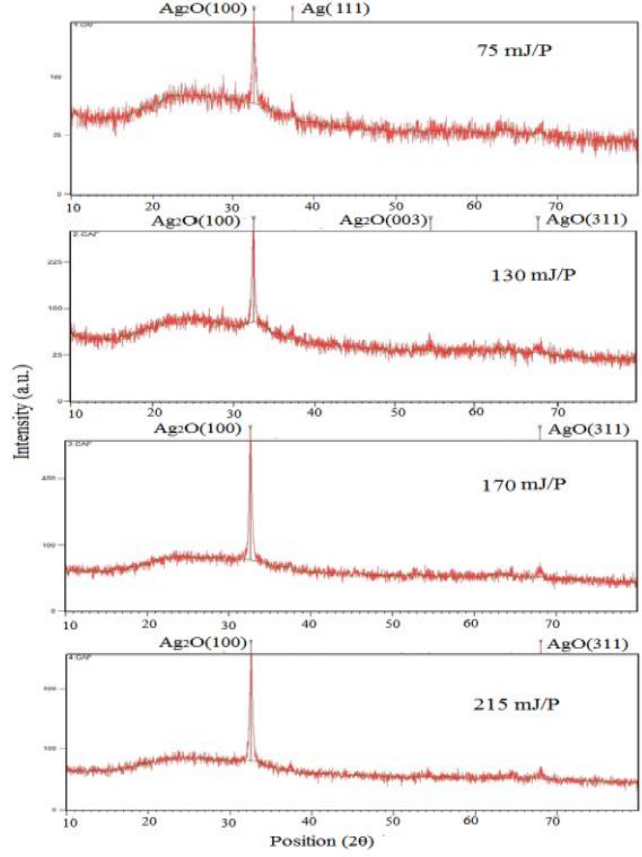

**FIGURE 1:** XRD Plot of the $Ag_2O$ films for different laser energy and identification of peaks.

The lattice constants of the crystal structures obtained from the XRD analysis are listed in **TABLE 1.** The crystallite size (l) of the grown films can be calculated by using the Debye-Scherrer formula [11]

$$l = \frac{0.9\lambda}{\beta cos\theta} \qquad (2)$$

where λ is the wavelength of X-rays, and β is the full width at the half maxima (FWHM) of the XRD peak at the Bragg diffraction angle θ. In our case, the crystalline size is measured for the most prominent peaks of the samples (100). Crystallite size stands for the dimension of the coherent diffracting domain. The smaller FWHM indicates the better crystallinity. One can estimate the dislocation density (δ) defined as the length of dislocation lines per unit volume, from the formula [12]

$$\delta = \frac{1}{l^2} \qquad (3)$$

**TABLE 1:** The values of interplanar distance, lattice parameters, crystallite size, dislocation density and strain calculated from the XRD analysis.

| Sample No. | Laser Energy (mJ/Pulse) | Interplanar Distance (d) (nm) | Lattice parameter (a,b) (nm) | Lattice parameter (c) (nm) | Crystallite size (l) (nm) | Dislocation density (δ)(nm$^{-2}$) | Strain (ε) |
|---|---|---|---|---|---|---|---|
| 1 | 75  | 0.27618 | 0.3189 | 0.5058 | 26.38 | 14.37×10$^{-4}$ | 13.14×10$^{-4}$ |
| 2 | 130 | 0.2767  | 0.3195 | 0.5062 | 26.62 | 14.11×10$^{-4}$ | 13.02×10$^{-4}$ |
| 3 | 170 | 0.27753 | 0.3205 | 0.5083 | 26.71 | 14.02×10$^{-4}$ | 12.98×10$^{-4}$ |
| 4 | 215 | 0.27912 | 0.3223 | 0.5112 | 27.27 | 13.45×10$^{-4}$ | 12.71×10$^{-4}$ |

The strain (ε) of the thin films is calculated from the relation

$$\varepsilon = \frac{1}{4}\beta\cos(\theta)\ldots\ldots(4)$$

The values of the crystalline size, dislocation density and strain are listed in **TABLE 1.**

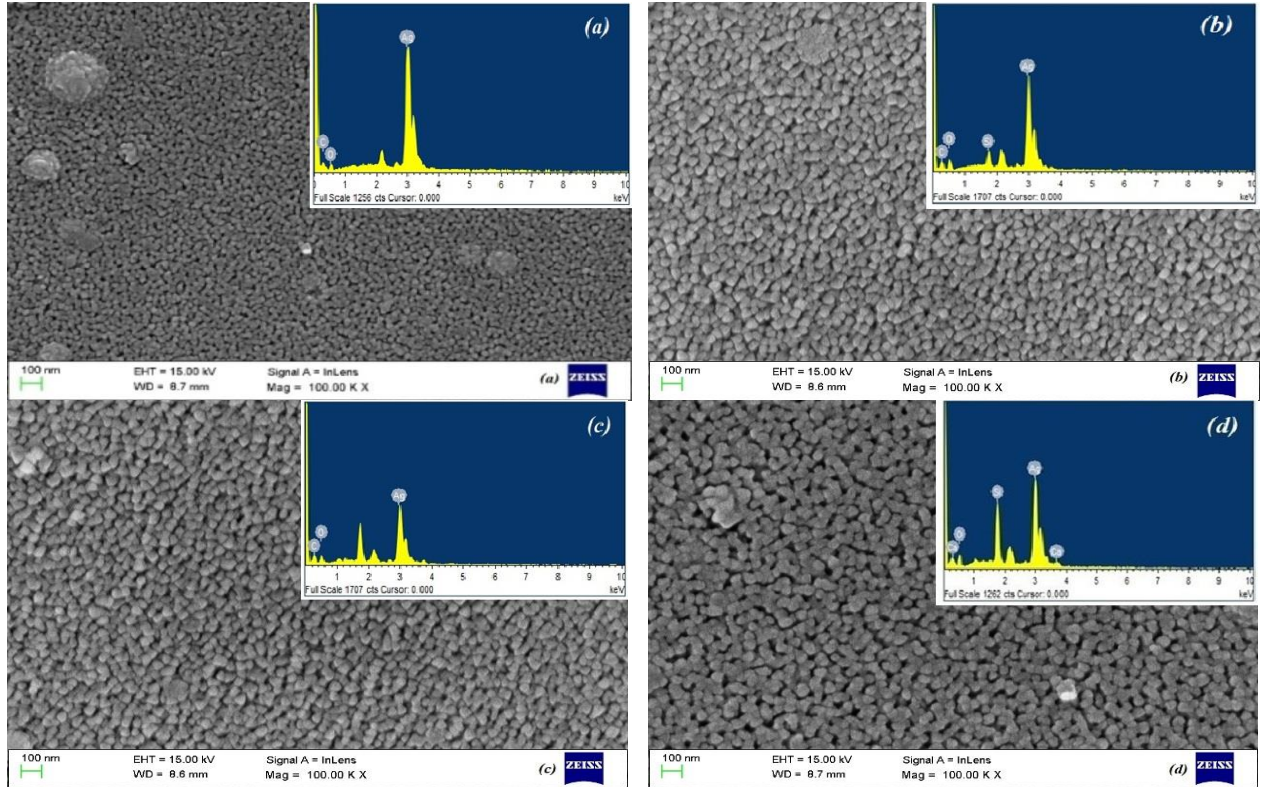

**FIGURE. 2:** FESEM images of the $Ag_2O$ films and the inset of the figures shows the EDX spectra for different laser energy. *(a)* 75 mJ/Pulse, *(b)* 130 mJ/Pulse, *(c)* 170 mJ/Pulse, *(d)* 215 mJ/Pulse

**TABLE 2:** Average diameter of nanoparticles obtained from FE-SEM Image, weight and atomic percentage of silver and oxygen obtained from EDX spectra. The optical band gaps of the films, calculated from the Tauc plot.

| Sample No. | Deposition Parameter (Laser Energy) (mJ/Pulse) | Diameter of the Particles (nm) | EDX result | | | | Optical Band Gap (eV) |
|---|---|---|---|---|---|---|---|
| | | | Weight% of Silver | Atomic% of Silver | Weight% of Oxygen | Atomic% of Oxygen | |
| 1 | 75 | 34.84 | 86.26 | 46.76 | 11.23 | 41.03 | 0.87 |
| 2 | 130 | 53.04 | 77.61 | 34.57 | 15.71 | 47.18 | 0.93 |
| 3 | 170 | 56.46 | 77.41 | 32.02 | 17.24 | 48.10 | 0.97 |
| 4 | 215 | 65.83 | 68.02 | 28.55 | 17.09 | 48.38 | 0.98 |

The FE-SEM images of the deposited films are shown in **FIGURE 2** which shows that the morphology of the films, although homogeneous, but are not totally uniform and the nanoparticles in the films are found as almost spherical shape. It is seen that the grain size increases gradually with the increment of laser energy and so on the grain boundary. The EDX spectrum, shown in the inset of **FIGURE 2**, shows that there is a raising of the percentage of the oxygen in the samples with the increment of laser energy. This might happen because of more oxidization of silver as the plasma ion contains more excitation energy with the increment of laser energy. Another reason can be trapping of more oxygen atoms inside the increasing of the grain boundary. In spite of silver and oxygen, it is seen that some other peaks also exist there which stand for the elements like carbon, calcium, silicon, etc.; are considered as impurity. Those might have come from the composition of glass substrate or impurity of target material. The average diameters of the particles for the four films, and the percentage of the oxygen and silver are listed in **TABLE 2.**

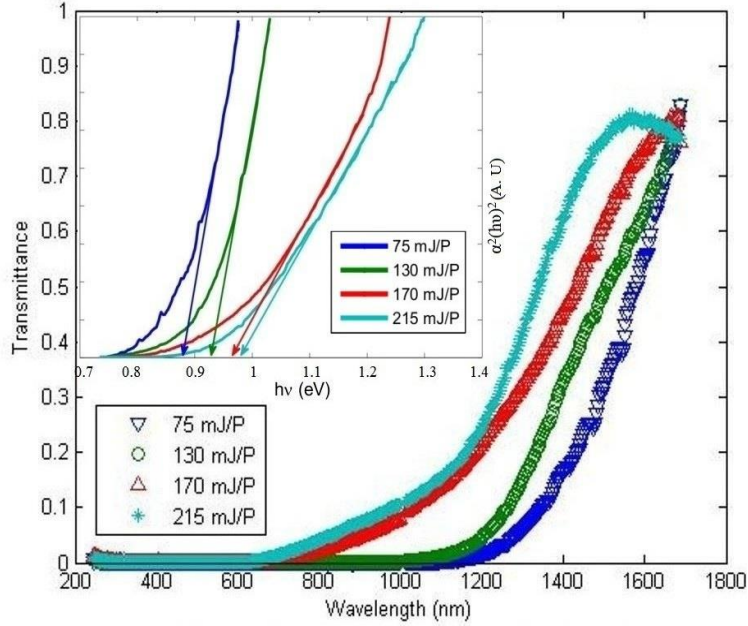

**FIGURE. 3:** Transmission spectra of the $Ag_2O$ films and Tauc's plot in the inset for different laser energy.

The transmission spectra, shown in **FIGURE 3**, indicates that the films are not transparent (transparency < 5%) in visible optical range, but becomes transparent in infrared optical range (transparency >70%). It is also seen that the cutoff wavelength varies for different samples. Note that it reduces with the increment of laser power. The reason might be the forming of better crystallinity with the increment of laser power and therefore, the achievement of more transmission. The band gaps of those samples are calculated from Tauc plot and listed in the **TABLE 2**. It is done by finding out the absorption coefficients($\alpha$) of the deposited films by using Lambert's formula [13]:

$$\alpha = \frac{1}{t} \ln \{\frac{1}{T}\} \tag{5}$$

Here, t is the thickness of the film and T is the transmittance. The optical band gap is calculated from the Davis and Mott equation as [13]

$$(\alpha h\nu)^2 = D (h\nu - E_g) \tag{6}$$

where D is a constant, h$\nu$ is the energy of a photon and $E_g$ is the optical band gap. The linear portion of the absorption edge of this plot has been extrapolated to find out intercept of the energy axis. From the Tauc plot, it is clear that the optical band gapes have appeared in the infrared range (energy of visible optical range is 1.6-3.2 eV). It is also notable that the optical band gapes increase with the increment of laser energy as well due to better formation of crystalinity. The red-shift might occur in the optical absorption spectrum due to the reduction of the particle size, which is known as the "quantum size effect"(QSE) [14]. The band gap is monitored by the density of states which is modified by the particle size of the nano grains and hence, the reduction of the particle size leads to have a blue shift in the band gap energy. Also, the reduction of band gap appears from electron-impurity and electron-electron scattering [15].

The optical reflection and absorption spectra are shown in **FIGURE 4(a)** and **4(b)** respectively. The deposited films have shown low optical reflection (reflectance<20%) and very high absorbance (> 80%) in visible optical range, which promises its applicability as a good absorbent in photovoltaic device application.

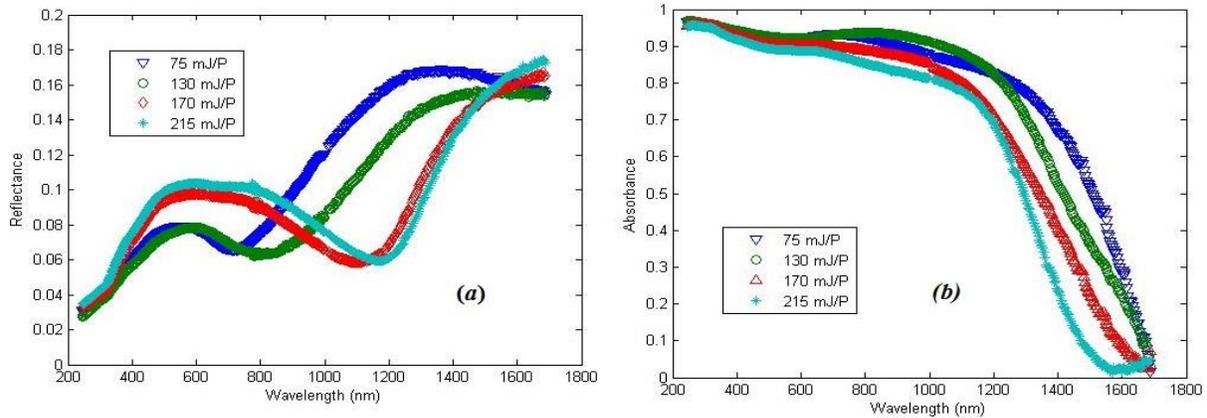

**FIGURE. 4:** (*a*) Reflection and (*b*) absorption spectra of the $Ag_2O$ films for different laser energy

## CONCLUSION

$Ag_2O$ films are deposited in PLD and after XRD, FESEM and optical characterization, it is confirmed that the films are formed as hexagonal crystal structure and having band gaps less than the energy of visible light. It is also seen that more deposition laser energy leads to get bigger grain size and more oxygen contamination in the sample and therefore, a blue shift is obtained in the band gap. For having p-type semiconductor behavior and good absorbance, we strongly recommend it to use as an absorbent for solar photovoltaic purposes.

## ACKNOWLEDGMENTS

I gratefully acknowledge the Ministry of Human Resource Development, India for their financial support.